\documentclass[manuscript]{acmart}
\AtBeginDocument{%
  }

\setcopyright{acmlicensed}
\copyrightyear{2026}
\acmYear{2026}
\acmDOI{XXXXXXX.XXXXXXX}
\acmConference[Conference acronym 'XX]{Make sure to enter the correct
  conference title from your rights confirmation email}{June 03--05,
  2026}{Woodstock, NY}
\acmISBN{978-1-4503-XXXX-X/2026/06}

\usepackage{array}
\usepackage{subcaption}
\newcolumntype{L}[1]{>{\raggedright\arraybackslash}p{#1}}

\usepackage{graphicx}
\usepackage{tikz}
\usetikzlibrary{shadows}
\begin{document}

\title{Imagining a Muslim Internet: Trust, Autonomy, and Segregation in a Faith-Aligned Browser}
%Porous Digital Segregation: How Muslim Users Navigate Trust and Autonomy in a Faith-Aligned Browser


\author{Umme Jannat Taposhi\textsuperscript{*}}
\affiliation{%
  \institution{BRAC University}
  \city{Dhaka}
  \country{Bangladesh}
}

\author{Farhan Tanvir Niloy}
\affiliation{%
  \institution{BRAC University}
  \city{Dhaka}
  \country{Bangladesh}
}

\author{Sabbir Bin Abdul Latif}
\affiliation{%
  \institution{University of Dhaka}
  \city{Dhaka}
  \country{Bangladesh}
}

\author{Farida Chowdhury}
\affiliation{%
  \institution{BRAC University}
  \city{Dhaka}
  \country{Bangladesh}
}

\author{S M Taiabul Haque\textsuperscript{\dag}}
\affiliation{%
  \institution{BRAC University}
  \city{Dhaka}
  \country{Bangladesh}
}

\renewcommand{\shortauthors}{Taposhi et al.}


\begin{abstract}
Religiously branded platforms raise important questions about trust, usability, and autonomy when technology is built around a specific faith. Prior HCI work on Islam and Muslim technology has focused on single-purpose tools such as prayer, scripture, and health apps, leaving infrastructures like browsers, which shape a user's relationship with the Internet, unexamined. We address this gap through semi-structured interviews with 16 users of Kahf Browser, a faith-aligned browser designed to support Muslims' online activities. Findings show religious identity motivates adoption but does not sustain it. Trust is not fixed by religious branding but shifts over time based on the browser's functionality, and users take pride in Muslim-built infrastructure. Building on these insights, we introduce porous digital segregation: a model where users seek not a sealed alternative internet but a protected default with selective exit mechanisms. We conclude by connecting findings to broader issues in faith-aligned technology and offering design recommendations.
\end{abstract}

% keyword: Faith-aligned Technology, Porous digital segregation, Internet fragmentation, Islamic HCI

%Participants overlook small technical issues, but when the core filtering falls short, it feels less like a bug and more like a broken promise, a line we call the usability-faith threshold.
\begin{CCSXML}
<ccs2012>
   <concept>
       <concept_id>10003120.10003121.10003126</concept_id>
       <concept_desc>Human-centered computing~HCI theory, concepts and models</concept_desc>
       <concept_significance>500</concept_significance>
       </concept>
   <concept>
       <concept_id>10003120.10003121.10011748</concept_id>
       <concept_desc>Human-centered computing~Empirical studies in HCI</concept_desc>
       <concept_significance>500</concept_significance>
       </concept>
 </ccs2012>
\end{CCSXML}

\ccsdesc[500]{Human-centered computing~HCI theory, concepts and models}
\ccsdesc[500]{Human-centered computing~Empirical studies in HCI}

\keywords{Faith-Aligned Technology, Trust, Digital Autonomy, Usability-faith Threshold, Porous Digital Segregation, Internet Fragmentation, Muslim Internet, Islamic HCI}


\maketitle

\section{Introduction}
Technology cannot be considered a neutral channel for human activities, as every interface incorporates assumptions about people’s behavior and encodes a specific ethical perspective in the artifact itself \cite{verbeek2006materializing}. This insight has led HCI to develop methodologies such as value-sensitive design that explicitly build technology around users’ values and not assume a general, value-neutral user \cite{friedman2013valuesensitive}. Consequently, HCI is taking inspiration not only from secular values but also from religious traditions to design ethical technologies, exploring the challenges that emerge when religious value are incorporated into design \cite{conwill2025challenges}. For many users, the religious values are not a preference that can be toggled on and off, but a framework that organizes everyday decisions, relationships, and time \cite{wyche2009extraordinary, rifat2022integratingreligion}. Although HCI has increasingly advocated to view religion as a mainstream design concern \cite{buie2013spirituality}, there are many areas where faith-based design are still at a nascent stage \cite{naqshbandi2022makingspace, rifat2022integratingreligion}, and research on religion and spirituality within HCI has remained relatively underdeveloped \cite{wolf2024stillnotalot}. As a result, secular design defaults have remained largely unexamined against religiously grounded daily life in many cases \cite{alsheikh2011whosevalue}, and web browsing technology is a prime example of this. 

Every leading digital platform is structured to capture and sustain users' attention, frequently through interfaces that manipulate rather than meet users' own objectives. This principle applies to browsers too; in fact, they are the delivery mechanism of the single layer that reaches nearly every other platform, but value-sensitive design has rarely been applied to the design of the browser itself, leaving its defaults implicitly secular by omission rather than choice. For an estimated 1.9 to 2 billion Muslims globally \cite{campbellesen2026islamicai, pew2025muslimpop}, this omission is not a design margin, but a daily friction. Mainstream recommendation systems routinely surface content that feels arbitrary rather than appropriate for Muslims \cite{jahan2026imaan}, because the values that are facilitating the recommendations are never theirs to begin with.

A growing body of research examines this kind of design choice (often referred to as Islamic HCI \cite{mustafa2020islamichci, ibtasam2021forgodsake}) where technology is built with Islamic values in mind from the outset as opposed to adapting them later, treating faith as a legitimate design consideration rather than a demographic variable \cite{noor2026decoding}. These works cover prayer apps, scripture apps, health platforms, hiring systems, and AI governance, spanning Muslim users negotiating trust with apps that mediate the Qur'an and Hadith \cite{easha2026islamiclearning} to Muslim users evaluating AI through Islamic concepts of honesty and accountability \cite{campbellesen2026islamicai}, algorithmic harms faced by Muslim job seekers on hiring platforms \cite{afnan2021asymmetries}, Eurocentric assumptions built into health platforms used by Muslim women \cite{alnaimi2024culturalreligious, yehia2026digitalmajlis}, and coordinated harassment targeting Muslim women online \cite{bhimdiwala2024fighting}. Religious values shape what technology Muslims trust and adopt \cite{alnaimi2024culturalreligious,yehia2026digitalmajlis} and drive their distrust of algorithms built in secular, Western contexts \cite{jahan2026imaan}, with that trust shaped by layers of disposition, situational context, and accumulated experience \cite{hoffbashir2015trustautomation}. None of this literature has examined the browser, the layer of daily digital life that sits upstream of every app, platform, and system it studies. In Bangladesh, positioned like much of the Global South as a consumer rather than a producer of digital infrastructure \cite{irani2010postcolonial, kumar2024digitalpostcolonial}, a group of developers set out to build one such value-based browser called Kahf \cite{kahf2026browser}, which blurs immodest images, screens gambling ads, and restricts un-Islamic contents by default. Kahf's objective echoes a longer debate over whether a single global Internet is giving way to culturally bounded alternatives \cite{mueller2017fragment}, positioning it as an early attempt at a faith-aligned corner of the Internet, yet the opportunities and challenges of a faith-aligned browser has remained unexplored in the HCI literature. 

We address this gap by examining how Muslim users come to trust, navigate, and imagine the future of Kahf -- a browser that is already built and in daily use. Prior work in HCI has explored religious values in individual apps and platforms but none has asked how users start using a browser built on these values, how they make sense of it once it is integrated into their daily lives, or how they envision its future. We do not assess Kahf as a system rather we look at the people who use it, where their trust originates, how they deal with its presence day to day, and what expectations they develop over time. This is important because Kahf exists at a stratum that earlier HCI studies on religious values have overlooked. It is not one religious feature added to a secular tool but an attempt to rebuild the tool itself, for a population whose relationship to the Internet has long been shaped by defaults it had no part in setting. In our study, we investigate the following research questions:
\begin{description}
    \item[RQ1:] How do Muslim users' religious values and identities shape their motivations for adopting and trusting Kahf, a faith-aligned browser?
    \item[RQ2:] How do users negotiate tensions between religious alignment, technological performance, and trust when evaluating Kahf?
    \item[RQ3:] How do users imagine the boundaries, governance, and future of a Muslim-aligned digital ecosystem, using Kahf as their point of reference?
\end{description}

To answer these questions, we conducted a qualitative interview study with 16 Kahf Browser users in Bangladesh. Instead of evaluating Kahf's filtering accuracy or interface design in isolation, we analyze participants' descriptions of adoption, trust, tolerance, and vision for Kahf, which, in turn, leads to a deeper understanding of a faith-aligned browser. Across these narratives, users express a preference for not being isolated from the larger Internet, but instead for a type of segregation that allows for a curated experience, filtering information without total disconnection.

Our analysis leads to three contributions. First, we show that faith is a starting point for trust, not an end point, and is continuously refined as participants trade off technical performance against religious alignment. Second, we introduce a usability-faith threshold, the point at which a technical flaw becomes perceived as a violation of religious commitment rather than a usability problem, and surface a related tension between pride in community-built technology and dependency on the platforms it positions itself against, extending existing work on digital autonomy in Muslim communities from single apps to infrastructure. Third, we ground the policy-level debate on Internet fragmentation \cite{mueller2017fragment} in real user data, introducing \textit{porous digital segregation} -- an Internet design model based on a protected default with a door in it rather than a sealed-off alternative, where users get a filtered experience by default but can still choose to step outside it. Finally, based on our findings, we present a few design implications.

\section{Background and Related Work}
This review draws on four bodies of work: Islamic HCI, halal and religious trust, technology acceptance and digital literacy, and Internet fragmentation. Together, they show how Muslim users evaluate religious claims and build trust in the products they use. What none of them has studied is an Internet system that claims to be religious in itself.

\subsection{Islamic HCI and the study of digital religion}

HCI has studied religious practice for two decades. Noor and Biswas \cite{noor2026decoding} recently reviewed HCI research on religion, spirituality, and faith and found that most of it documents how Muslims use general-purpose technology, rather than examine systems designed from Islamic principles. Barzilai-Nahon and Barzilai \cite{barzilai2005cultured} supply a name for the alternative: cultured technology, the process by which a religious community adopts a technology and reshapes it to fit its own values. Their case -- ultra-Orthodox Jews who continued to go online despite an explicit rabbinical ban on Internet use -- illustrates that it is the norms and social structures of the community, rather than the technology itself.

Later studies extend this picture to other religious contexts: how sermons are prepared and delivered with the help of technology \cite{rifat2022waz, rifat2022situating}, how mosques and madrasahs sustain everyday religious practice over time \cite{rifat2020religionsustainability}, how ritual observance was adapted and maintained during the COVID-19 pandemic \cite{caidi2023ramadan}, and how people engage in spiritual reading through Islamic websites \cite{bachtiar2022capturing}. Jahan et al. \cite{jahan2026imaan} come closest to our study, finding that Bangladeshi Muslims treat recommendation algorithms as a spiritual guide where users valued a curated ``halal feed'' but worried how much of their religious life the algorithm now controlled, and called for Islamic scholars to enter the loop. Their objective, though, was a set of features within apps like YouTube and Facebook, not a standalone Islamic product, and no study has asked how a user would trust or distrust that a complete faith-aligned infrastructure in the first place. That gap motivates us to ask why Muslim users adopt and trust a browser built to be religious by design (RQ1).

\subsection{Halal as a credence claim: trust, verification, and acceptance models}

According to Islamic law, ``halal'' is an arabic word that means ``permissible'' or ``lawful''. The concept of Halal has been explored in different aspects, including mortgage \cite{calder2020halalization}, food and consumer products \cite{fischer2011halal}, and organizational practices \cite{adham2024halal}, among others. Collectively, these works define halal as a credence attribute, where the trust or acceptance comes from two sources: (i) formal certification of a halal product/service, or (ii) personal trust in another Muslim vendor with a good reputation \cite{bonne2008trust}. These ideas are subsequently extended into the software industry, with a notable example being SalamWeb -- a Shariah-compliant browser certified by Amanie International Shariah Supervisory Board, following Malaysia's Islamic Digital Economy standard \cite{ramli2022salamweb}. Kahf's judgment, in contrast, comes from an in-house Mufti team \cite{kahf2024futurestartup}, and it seems reasonable to examine how this credential translates into user confidence and trust.  

Bangladeshi religious media shows how such credentials work informally. Rifat et al. \cite{rifat2022waz} describe media sustained by personal commitment, not institutional backing, while Raya \cite{raya2025digital} describes Indonesian preachers turning religious authority into business via apps and social media. Stated trust does not always predict behavior. Alam et al. \cite{alam2025blind} found that Muslim participants who said they trusted scholars over AI chose the AI's answers once the comparison was blind. Removing a certifier does not remove the underlying judgment, it only relocates it. A blocklist still encodes someone's view of what a Muslim should not see, whether that someone is a scholar, a board, a development team, or an assumed consensus nobody actually confirmed \cite{raya2025digital, rifat2022situating}. Where no board has ruled, users judge the filter themselves, applying the same checks they would to a sermon: which school of thought it follows, what political leanings it carries.

That capacity is unevenly distributed. Latif et al. \cite{latif2026paradox} found digital literacy determined whether Bangladeshi citizens completed an e-government task alone or needed a mediator, and that mediated access helped even as it created dependency. A faith-aligned browser poses the same problem in miniature: its claims are checkable only by someone who can read permission settings or judge what a missing code repository implies, a gap Noor and Biswas \cite{noor2026decoding} say Islamic HCI has largely ignored. Unable to verify, users theorize instead, building an account of a filter's hidden workings from whatever they can see. Eslami et al. \cite{eslami2015always} found most people did not know Facebook's News Feed was curated at all, and that discovering it changed how they read their own friendships. DeVito et al. \cite{devito2018folk} trace how such folk theories form from scattered clues, and Bucher \cite{bucher2017imaginary} argues the theories loop back to shape what the systems become \cite{devito2021adaptive}. More detailed theories produce more specific reactions to change \cite{devito2017algorithms}, and a halal label sets an unusually clear expectation against which any change can be measured. Razaq and Ghoshal \cite{razaq2024muslim} found Muslim users who believe the network is watching them and limit their own searches accordingly, a fear with a documented basis since location data harvested from a prayer app reached US military buyers through a data broker \cite{muslimpro2020vice}. Software offers a further check food cannot, open-source code lets anyone inspect how a system works, a third option beyond the two Bonne and Verbeke describe. Whether users treat this as a religious requirement or merely a convenience remains unasked.

The same pattern surfaces from the opposite direction in survey research on religious branding, most of it built on the Technology Acceptance Model \cite{davis1989perceived, venkatesh2000theoretical}. Usman and Projo \cite{usman2022integration} surveyed 300 Islamic banking customers and found that adding sharia compliance and trust raised loyalty and the same pattern recurs with religiosity in Indonesian small businesses \cite{setiawan2024religiosity}, Muslim students using AI for religious study \cite{faizin2025muslim}, and credibility and religious experience in a Catholic prayer app \cite{haryanto2024modifying}. In each case, more of the religious factor predicts more acceptance, a gap Bagozzi \cite{bagozzi2007legacy} already names in TAM more broadly but the model has little to say about how that acceptance is reappraised once the technology is actually in use.

Research on brand trust complicates this picture. Trust builds slowly and breaks fast \cite{slovic1993perceived}. Aaker et al. \cite{aaker2004goodbrands} found mistakes damaged sincere-brand relationships but revived exciting-brand ones, and Gr{\'e}goire et al. \cite{gregoire2009love} describe a love-becomes-hate effect: a brand's most loyal customers hold the longest grudges after failure. Acceptance models also stop too early, ending at first use, so a user who installs a faith-based app and later deletes it counts as no data at all, even though HCI has long treated non-use as informative rather than absent \cite{satchell2009beyond, baumer2011implication}. None of this makes the acceptance literature wrong, only narrow where it answers the seller's question, not the user's, since ease of use and a legible religious signal are exactly what a company can control.

None of these studies capture the negotiation that happens after adoption, how a user weighs a technical flaw against the value alignment that originally drew them in, and at what point that weighing tips into distrust. This post-adoption negotiation is what our second question examines (RQ2).

\subsection{Internet fragmentation and the boundaries of a Muslim Internet}

People have worried about the Internet splitting apart for some time, under names that track the cause as splinternet, cyber-balkanization, digital sovereignty \cite{mueller2017fragment}. China's Great Firewall and Russia's Sovereign Internet Law are the clearest cases, both cutting a population off from the open Internet for political reasons. DeNardis \cite{denardis2009protocol} documents a quieter, more technical version, a shortage of IPv4 addresses pushed countries toward incompatible systems. Mueller \cite{mueller2017fragment} pushes back: a fully broken Internet is unlikely, since network effects make a shared global system too valuable to abandon, and what is really happening is a fight over sovereignty, of which fragmentation is a side effect rather than the goal. The pressure is not confined to authoritarian states: Brazil, Germany, and the UK have all pushed to bring parts of the Internet under their own rules.

This debate centers on governments and rarely asks about communities that choose enclosure for themselves. Barzilai-Nahon and Barzilai \cite{barzilai2005cultured} offer an opening to describe seclusion as one way a religious community reshapes technology to fit its own values, rather than having enclosure imposed from outside. A related idea already exists for Muslims specifically where researchers describe a digital or virtual ummah, an online space where Muslims scattered worldwide build a shared religious community that mirrors the offline one \cite{khamis2021cyberummah}. But this research treats the digital ummah as a social space inside the regular Internet but it does not ask what happens when a product tries to wall that space off technically, the way a government would. Razaq and Ghoshal's borders \cite{razaq2024muslim}, by contrast, are imposed on Muslim users from outside, closer to state-driven fragmentation than to a boundary a community sets for itself. Neither their work nor the digital ummah literature distinguishes a self-chosen boundary from one imposed by a government or company.

What is missing is any account of how people inside a bounded space judge it: whether it feels like protection or confinement, whether they can leave, whether they chose to enter, and who decides who else is let in. Our third question asks this directly of Kahf's users, not policy analysts (RQ3).

\subsection{Kahf: a browser built for religious compliance}

\begin{figure*}[t]
    \centering

    \begin{subfigure}[t]{0.49\textwidth}
        \centering
        \includegraphics[width=\linewidth]{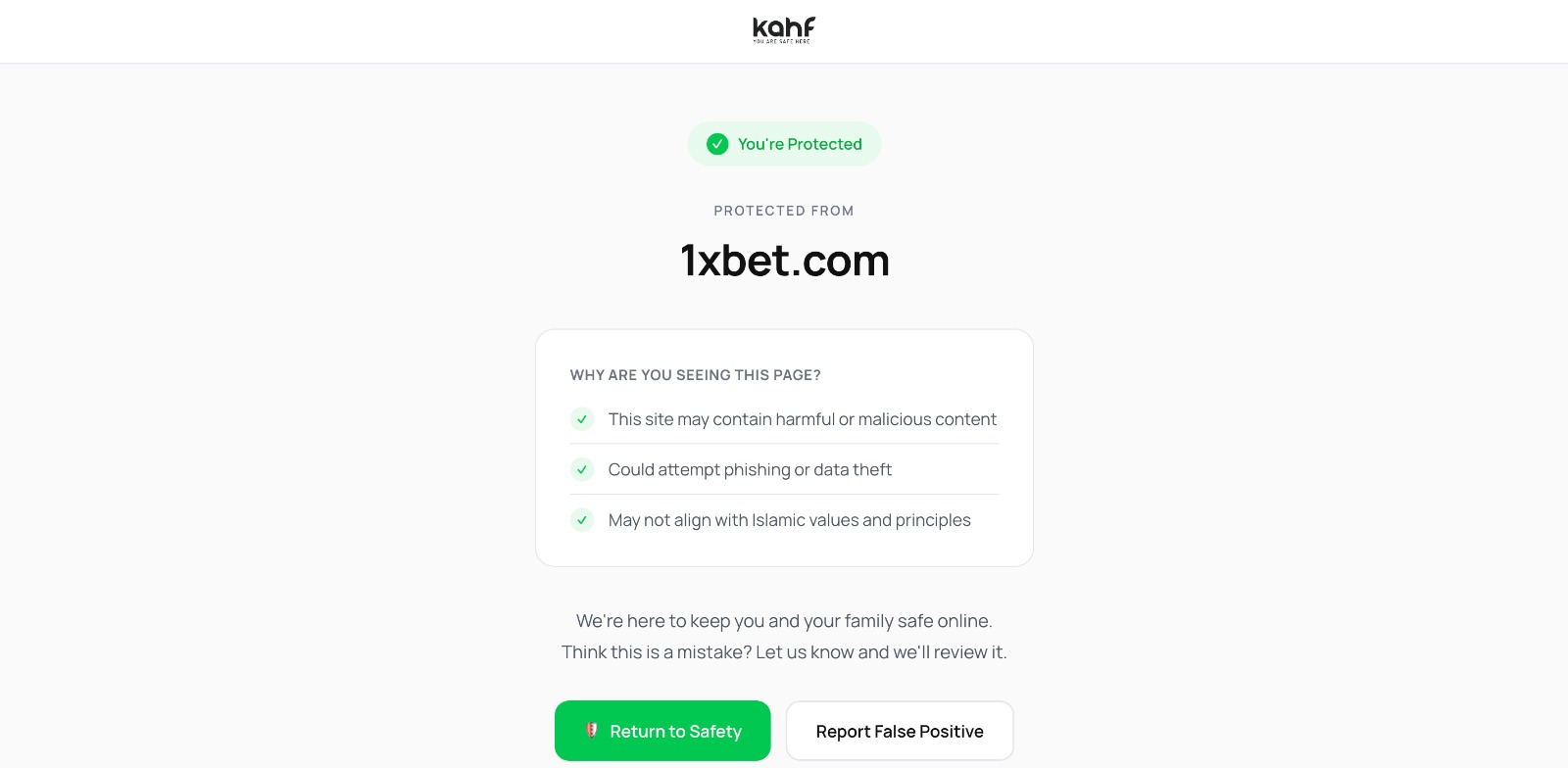}
        \caption{Domain blocking for a gambling website.}
        \label{fig:kahf-blocking}
    \end{subfigure}
    \hfill
    \begin{subfigure}[t]{0.49\textwidth}
        \centering
        \includegraphics[width=\linewidth]{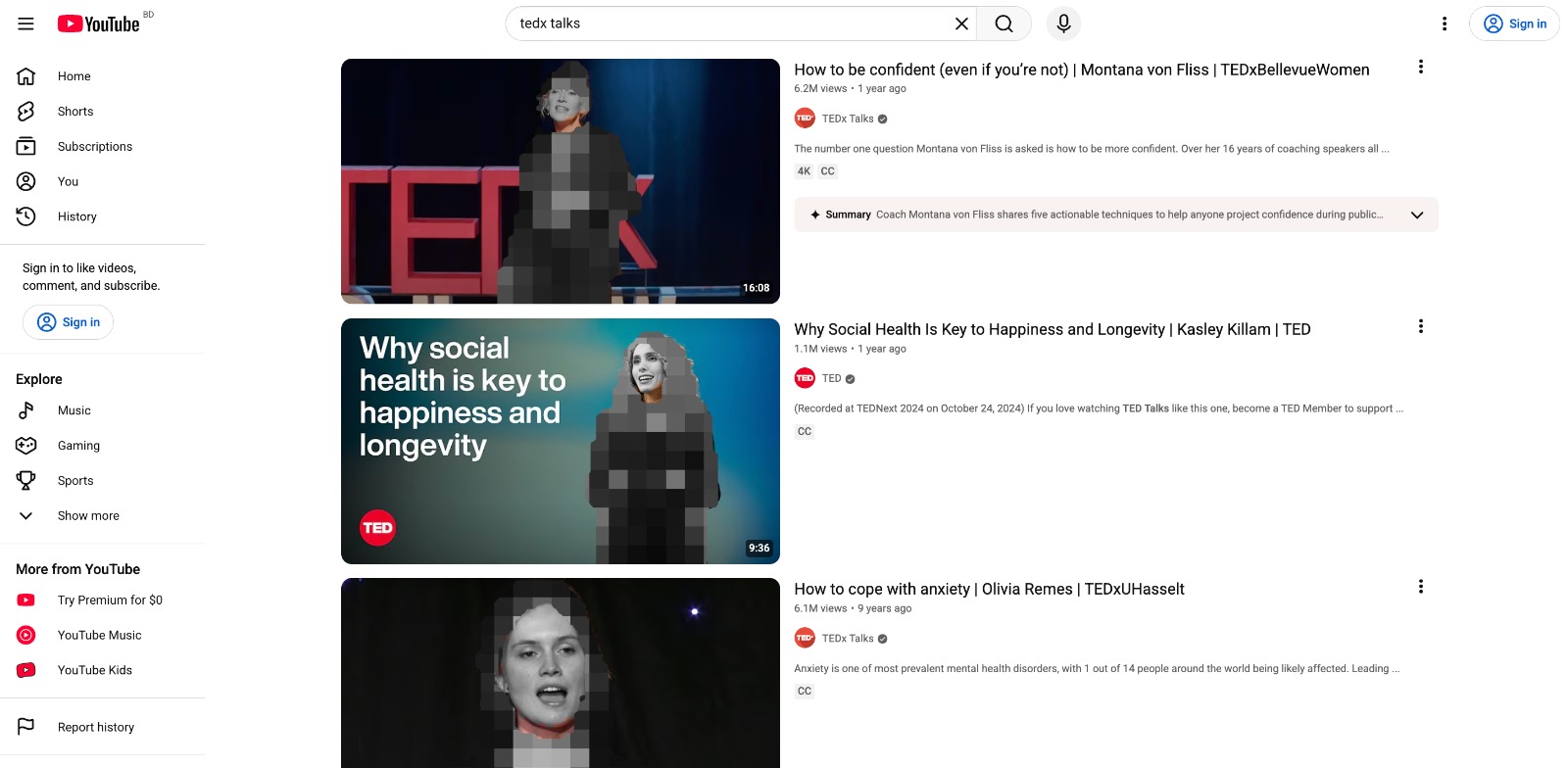}
        \caption{Automatic blurring of detected visual content.}
        \label{fig:kahf-blurring}
    \end{subfigure}

    \caption{Examples of Kahf Browser's content filtering: 
    (a) a domain-blocking page shown when a restricted website is accessed, and 
    (b) automatic blurring of detected visual content in YouTube search results while other interface elements remain accessible.}

    \Description{Two screenshots demonstrating Kahf Browser's filtering mechanisms. The left screenshot shows a protection page blocking access to a restricted website (gambling website) and explaining that the site may contain harmful content or content that does not align with Islamic values. The right screenshot shows YouTube search results in which portions of people appearing in video thumbnails are automatically pixelated while the surrounding interface, titles, and other information remain visible.}

    \label{fig:kahf-filtering}
\end{figure*}

Kahf began as a startup founded in 2022 by a Bangladeshi entrepreneur \cite{kahf2025islamonline}, first covered under the name Halalz. Its earliest product, Kahf Guard, blocked haram content at the network level \cite{kahf2024futurestartup, kahf2024dailystar}. The browser now operates under the corporate name Kahf Yaz{\i}l{\i}m A.{\c S}., registered in T{\"u}rkiye, and reports over 280,000 active installs and more than 1.6 million blocked unsafe sites, with a 4.5-star rating on Google Play; the app is free, requires no account, and runs on Windows, macOS, iOS, and Android \cite{kahf2026browser}.

Five protections come enabled by default: an AI image classifier that blurs immodest visuals as a page loads, a safe search mode, a restricted YouTube mode, an ad and tracker blocker that removes gambling and adult content, and a continuously updated blocklist for haram sites, phishing, and malware \cite{kahf2026browser}. A Bangladeshi Islamic scholar and a professor from King Saud University both publicly vouch for the browser \cite{kahf2026browser}, and the company's other products form, together with Kahf, the ``halal Internet ecosystem'' this work studies.

~\hyperref[fig:kahf-filtering]{Figure 1} illustrates two of Kahf's filtering mechanisms in practice: blocking access to restricted domains and selectively blurring detected visual content while leaving the surrounding webpage accessible.

\section{Methodology}
We conducted a qualitative interview study with 16 Kahf Browser users. The interviews explored participants' experiences with technology use, religious values, trust, and faith-aligned computing. \autoref{tab:participants-users} summarizes participant demographics and experiences with Kahf Browser.

% ---------- Table 1: Kahf Browser Users ----------
\begin{table*}[t]
\centering
\caption{Participant demographics and Kahf usage experience.}
\label{tab:participants-users}
\renewcommand{\arraystretch}{1.25}
\setlength{\tabcolsep}{4pt}
\footnotesize
\begin{tabular}{@{}
    L{0.05\textwidth}
    L{0.08\textwidth}
    L{0.10\textwidth}
    L{0.18\textwidth}
    L{0.18\textwidth}
    L{0.09\textwidth}
    L{0.10\textwidth}
@{}}
\toprule
\textbf{ID} & \textbf{Gender} & \textbf{Kahf Experience} & \textbf{Education} & \textbf{Profession} & \textbf{City} & \textbf{Recruitment Channel} \\
\midrule
P01 & Male   & 2 years     & BSc, Mathematics      & Student           & Nilphamari & FB \\
P02 & Male   & 1.5 years   & Pass Course (Degree)  & Student           & Narsingdi  & FB \\
P03 & Female & 1 month     & BSc, Mathematics      & Student           & Sylhet     & UNI \\
P04 & Female & 2 months    & M.A, Social Work      & Student           & Sylhet     & UNI \\
P05 & Male   & 6--7 months & BSc, CS               & Software Engineer & Dhaka      & UNI \\
P06 & Male   & 6--7 months & BSc, CS               & Recent Graduate   & Dhaka      & FB \\
P07 & Male   & 1.5 months  & BSc, CSE              & Student           & Dhaka      & UNI \\
P08 & Female & 3-4 moths   & BSc, CS               & Student           & Dhaka      & UNI \\
P09 & Female & 6--7 months & Pass Course (Degree)  & Student           & Cumilla    & FB \\
P10 & Male   & 6 months    & Msc, CSE              & Service Holder    & Jessore    & FB \\
P11 & Male   & 3 years     & not reported          & Service Holder    & Dhaka      & FB \\
P12 & Female & 2 months    & BSc, CSE              & Student           & Dhaka      & UNI \\
P13 & Male   & 4 months    & BSc, CSE              & Student           & Dhaka      & FB \\
P14 & Male   & 2 months    & Not reported          & Student           & Dhaka      & FB \\
P15 & Male   & 2 months    & Not reported          & Service Holder    & Khulna     & FB \\
P16 & Male   & 9 months    & BSc, CSE              & Student           & Dhaka      & FB \\
\bottomrule
\end{tabular}
\vspace{3pt}

\parbox{\textwidth}{\raggedright\footnotesize \textbf{Channel:} FB = direct outreach through Messenger from the Facebook Kahf Browser Groups; UNI = University community.}
\end{table*}

\subsection{Participants Recruitment}

We recruited a total of 16 participants for the study by using purposive sampling to recruit participants with experience using Kahf Browser. Recruitment took place through three channels.

\begin{itemize}
    \item \textbf{Facebook group posts.} We posted recruitment advertisements in three Facebook groups directed to Kahf Browser users, including a fan community for the browser. The posts described the research and offered BDT 500 as compensation.\footnote{Approximately USD 4 at the time of the study.} These generated limited interest, and the post in the Kahf-specific group did not yield any participants.

    \item \textbf{Direct outreach.} We reviewed discussions in the comment sections about Kahf Browser on the Kahf-specific Facebook groups manually. We found participants who made posts or comments indicating experience with Kahf Browser and reached out via Messenger. Many did not reply, and some refused to participate. Most of our sample was produced by this outreach.
    
    \item \textbf{University community.} We circulated the recruitment call with the university community, which helped in recruiting the rest of our additional participants.
\end{itemize}

We chose individual interviews because the personal nature of the study involved browsing practices, values, trust, privacy, and experience with technology aligned with faith. Prior HCI research states that group discussions can help create a blend of experiences, but individual interviews provide greater privacy when discussing personal or sensitive topics \cite{jahan2026imaan}. A one-to-one setting allowed participants to discuss their experiences and perspectives without the presence or influence of other community members. 

\subsection{Data Collection}

We conducted semi-structured interviews to gather data from the Kahf Browser users. At first, a pilot interview was conducted with a user whom we got from direct outreach, to refine the interview protocol. Based on the pilot interview, we revised questions that appeared unclear or potentially leading before conducting the remaining interviews. The authors conducted the interviews between June and August 2026 via Messenger calls, WhatsApp calls, and Google Meet. Interviews lasted approximately 42 minutes on average. All the interviews were primarily conducted in Bangla. We audio-recorded each interview with participants' consent. 

Our interview protocol covered the following topics: demographic background, religion and digital life, experience with Kahf Browser, trust and privacy, and views on a separate or bounded Muslim digital environment. Through these questions, we examined technology adoption, religious values, source credibility, risk perception, moral agency, in-group trust, and digital boundaries.

\subsection{Data Analysis}

For data analysis, the first two co-authors translated and transcribed the audio recordings manually. Analysis started alongside data collection, providing a context-based understanding that informed subsequent interviews. We used thematic analysis~\cite{braun2006using} and an inductive, open-coding approach~\cite{corbin2008basics}. The first two co-authors coded each transcript individually, labeling data segments with their own codes. The labels each author produced were then compared. When labels matched, they were accepted as consistent, and disagreements were resolved through negotiated agreement among the research team. Example codes include ``moral agency'', ``in-group trust identity'', ``usability tradeoff accepted for religious value'', ``inconsistent blur tool performance'', ``privacy distrust in mainstream platforms'', ``explicit support for segregated Muslim Internet'', and ``bounded openness in Islamic platform vision''. We did not use predetermined themes to maintain openness to emerging patterns in the data. The whole team then worked together to group related codes into broader themes that best illustrate our findings.

\subsection{Positionality}

All the co-authors of this work are Muslims. However, this study was shaped by our perspectives as researchers as well as our scholarly experiences. The study was conducted in the same country and language as our participants, allowing us to pick up on local expressions while participants spoke in their language of comfort. None of the authors were affiliated with the Kahf Browser team or personally acquainted with any participants. Since Kahf serves a fairly niche and conservative community, some effort was made to build rapport with our participants. 

\subsection{Ethical Consideration}
This study was approved by the university's Institutional Review Board, and informed consent was obtained from all participants, who took part entirely voluntarily. We initially attempted to recruit participants via public Facebook posts offering small compensation, but this yielded no responses; we then shifted to direct outreach and university community, where participants voluntarily shared professional insights without any financial compensation for their time. No participant was pressured or obligated to take part. All data were anonymized, with participants identified as P1–P16, and any administrative figure directly mentioned by participants referred to as the `a public face of Kahf'.

\section{Findings}
Our analysis revealed five themes across the sixteen participants (P01-P16), spanning participants made use of Kahf as moral infrastructure, how participants' trust shifted between faith and verification, how participants weighed usability versus religious alignment, how pride in the product coexisted with awareness of dependencies, and how participants imagined the frontiers of a Muslim Internet. ~\hyperref[tab:glossary]{Table 2} defines the Islamic terms used in this section.

\begin{table*}[t]
\caption{Glossary of Islamic terms used in this section.}
\label{tab:glossary}
\begin{tabular}{p{0.12\textwidth} p{0.82\textwidth}}
\toprule
\textbf{Term} & \textbf{Definition} \\
\midrule
Halal & Permissible according to Islamic law. \\
\addlinespace
Haram & Prohibited according to Islamic law \\
\addlinespace
Taqwa & God-consciousness. A person's innate awareness of God that influences daily behavior. \\
\addlinespace
Ummah & The global community of Muslims, understood as a shared identity that extends across national and cultural boundaries. \\
\addlinespace
Ulama & Islamic scholars recognized within Muslim communities for their religious knowledge and authority. \\
\addlinespace
Wasilah & A means through which a goal is achieved, often religious and spiritual. \\
\addlinespace
Quran & The central religious scripture of Islam which is believed by Muslims to be the word of God. \\
\addlinespace
Fitna & A trial or test of faith, it can also be used of social discord, strife or temptation that threatens moral or communal order. \\
\addlinespace
Alhamdulillah & An Arabic phrase meaning ``praise and thanks be to God,'' commonly used by Muslims to express gratitude. \\

\bottomrule
\end{tabular}
\end{table*}

\subsection{Technology as Moral Infrastructure: Guiding the Self and Others}
\subsubsection{Technology as a Supplement to Personal Discipline}

Participants described their online conduct as an integral part of their broader religious life, not a distinct space with permissive rules. This \textit{religious conduct as digital conduct} meant that the same standards that participants applied to eating, dressing or speaking offline were carried over into browsing, searching, and consuming content online, even though nobody directing them to do so. P04 expressed this most directly:

\begin{quote}
\textit{``As a Muslim, I follow a code of conduct up to a point. It is our way of life, according to Islam. There are regulations regarding halal and haram. I try to follow these rules in the way I move, eat, sleep, wake up and interact with people. I try to live by this same code of conduct in my digital life too.''} (P04)
\end{quote}

Participants stated that it was harder to stay disciplined online than offline, mostly because the unwanted content was presented to them without  being sought, through uncontrollable mechanisms rather than intentional search. This distinction separates two problems a protective browser may address: i) helping a user to avoid content they are tempted to seek, ii) protecting them from content they have never requested for. Several participants mentioned advertising as the main source of the second problem, P05 reported the surfacing of gambling ads and other objectionable content even during ordinary browsing. Such advertisements appear in almost every app and webpage which make him uncomfortable.

Given this consistent unsolicited exposure, no participant regarded Kahf as a comprehensive solution rather, it was regarded as a complement to self-discipline. P05 was most explicit about this limit, providing a rough ratio that put majority of the moral responsibility back on himself:

\begin{quote}
\textit{``It will not eliminate the need for self-control completely. In terms of ratio, if I put it, maybe 10\% comes from the tool and I have to do the rest 90\% myself. This doesn’t make it a total solution to staying good. If 10\% is coming from the tool, the other 90\% has to come from my own sincere intention, my own mindset.''} (P05)
\end{quote}

P07 came to a similar conclusion but with a different framing. According to him, technology and taqwa are complementary rather than competing:

\begin{quote}
\textit{``Both are needed, taqwa and technology. If a tool helps me with my taqwa, or helps me stay away from haram content, I think it should definitely be used. To prevent the misuse of technology, we definitely need technology.''} (P07)
\end{quote} 

P10 reached a similar view through a more concrete metaphor, describing taqwa being already inside him, with the browser simply eliminating the chance to act against it: \textit{``Taqwa is an internal, deep matter. This has to exist. If it does not exist, you will fall into trouble, right?''} Not every participant relied on Kahf's automated filtering for this discipline. P13 achieved a similar effect through his own behavior, accessing Facebook only through Kahf's less polished interface, since this friction discouraged him from spending long time on the platform, working alongside the browser's filtering rather than depending on it.

\subsubsection{Protecting Family Members on Shared Devices}

Beyond individual discipline, participants described a \textit{shared-device responsibility} extending their concern to family members using the same devices. Most participants used Kahf as their default browsing app on their devices for accessing sites like YouTube and Facebook rather than through a dedicated app. Since everyday browsing runs through Kahf by default, its filtering covers anyone who uses that phone or other devices, not just its owner. P02 provided the clearest example:

\begin{quote}
\textit{``The issue arises within my family. I am the elder brother, I have a younger brother, then there are my parents. Sometimes they use my mobile for different purposes and they may see bad content which can make them uncomfortable.''} (P02)
\end{quote}

P10 expressed a similar concern about the constant, low-level intrusion of unwanted material through channels other than browsers, such as messaging apps like IMO. He said he found such encounters disturbing and had taken to warning or blocking those responsible. At one point he deleted IMO altogether before he stumbled across Kahf. Kahf’s own filtering is but one aspect of the shared-device concern that his account brings into the larger digital environment. Whereas P08 adopted Kahf specifically to protect her teenage nieces and nephews on her devices, at her sister’s recommendation, P10 extended this logic further outward, framing the browser’s filtering as protection for the wider \textit{Muslim Ummah}.

P11 provided the most organized explanation of what the browser actually filters, linking these categories to religious identity as their justification:

\begin{quote}
\textit{``It ensures that four types of things do not get to the user. First of all, any kind of adult content. Secondly, gambling or betting. Third, malware and phishing. Fourth, this is a halal browser for the Muslim Ummah or Muslim community, and it blocks any anti-Islamic or Islamophobic content.''} (P11)
\end{quote}

He used the same language others did to talk about personal self-restraint, protecting his taqwa and his gaze, and broadened the scope of this filtering to younger users, framing the browser as protection for an entire generation:

\begin{quote}
\textit{``Our children, our Gen Z, our Alpha, our brightest generation are entering a dangerous digital space, a digital Babel. We must do everything possible to get them out of this distraction. I think Kahf Browser is going to be one of the front-runners who have taken this kind of initiative, what we call a wasilah, a means, to be saved from this distraction.''} (P11)
\end{quote}

Personal restraint and family protection emerge as two sides of the same underlying commitment, one aimed inward and other outward.

\subsection{From Faith-Based Trust to Verifiable Trust}

\subsubsection{Identity as a Starting Point for Trust}

For many participants, religious branding alone built an initial baseline of trust in Kahf, an \textit{identity as an initial trust signal} pattern that required no prior knowledge of the browser's technical features, resting instead on a general association between Islamic branding and good intent. P10 described this reaction as almost unavoidable, saying any Islamic symbol or logo gives him a sense that something good might be there. P02's account illustrates a related version, backed up by scholars:

\begin{quote}
\textit{``Many of the large ulama in the country have vouched for this, that it is a good browser. Because of them, I mostly downloaded the browser. At this point, I do not have full knowledge of how they work, nor have I really tried to find out.''} (P02)
\end{quote}

He expected the same scholars to flag any future problem, since as an ordinary person he would have no way of knowing if developers had shared user data with non-Muslim apps or organizations. A related but different pattern is observed in P12's account, who downloaded and began using the browser without any research, and only investigated the background afterward, deepening her trust only after this unplanned check.

Meanwhile, several participants viewed identity as insufficient evidence of trustworthiness, arguing that a religious label describes intention, not necessarily assurance. P03 extended the same suspicion to all branding:

\begin{quote}
\textit{``Just because it has Quran recitation doesn't mean there isn't some other line in there as well. I have my doubts about that. I don't trust it completely, whether it has an Islamic or non-Islamic name. I am always skeptical when it comes to trust.''} (P03)
\end{quote}

P04 pushed this further, claiming that religious branding acts as a liability, lowering a user’s guard precisely because the product appears trustworthy:

\begin{quote}
\textit{``Not everything is trustworthy. I am a believer in Islam without any question of doubt. But in the case of the person who follows Islam, that person cannot be trusted blindly. Everything we use, be it software or a browser, is made by a human. Whatever their intentions, it is best not to blindly trust these things.''} (P04)
\end{quote}

P06 did not doubt the Islamic label itself, but rather saw two types of threat to an Islamic label: people who openly undermine Islam from outside the religion, whom he sees as easy to identify, and people who present as Muslim while undermining the same interests from within, whom he sees as much harder to filter out. His logic suggests that skepticism about Islamic branding was not a rejection of religious identity as a signal, but an acknowledgment that the same identity could be claimed with very different intentions.

\subsubsection{Verification and Conditional Trust Over Time}

A separate group investigated the browser's background before installing, rather than checking afterward. P05 described a specific sequence of checks:

\begin{quote}
\textit{``I did not just downloaded it out of the blue. First, I checked who is making this, which country it is coming from, and who is on the developer team. Once I felt this could be trusted, I went to the Play Store and installed it.''} (P05)
\end{quote}

Even so, he did not treat verification as replacing religious branding entirely, noting an Islamic name still creates a sense of trust alongside his own research. P06 practiced a narrower, more immediate version, checking a product's privacy policy directly before downloading.

He also argued that trust should rest on evidence any user can independently check, and proposed open-sourcing the browser as the way to make that possible. This matters because he applied the same scrutiny to Kahf that he applied to any other product, rather than trusting it more simply because it shared his religious identity. Some participants extended trust to the organization's motives even without checking its code. P13 described Kahf as relying heavily on faith rather than commercial calculation, framing its stated goal of keeping Muslims' resources within Muslim hands as evidence of good intent, even where he could not verify it technically. His own account complicates this picture further, despite his technical training as a computer science student, he admitted not knowing how his data was being handled, saying it is not really possible for him to know. Though from a customer-experience perspective he is reasonably satisfied and does not verify apps beforehand absent negative reports.

P11 offered the most concrete example of independently verifiable evidence functioning as an actual trust mechanism:

\begin{quote}
\textit{``There is no email or ID authentication needed to use Kahf Browser, and I believe Kahf Browser has no backend at all, so tracking a user's history from the server end is not possible. They also introduced a `clear on exit' option, so users can clear whatever is stored locally as well.''} (P11)
\end{quote}

P15 raised a related concern around payment data, explaining that shopping sites often cache sensitive information like card numbers, CVVs, and expiry dates in the browser; he trusted Kahf specifically because, unlike others, it does not collect this data or keeps it inaccessible even when stored locally.

Across the sample, trust in Kahf proved \textit{conditional and revocable} rather than fixed. P05 said a confirmed data breach would be sufficient to make him abandon the browser, though a visible correction could restore his trust afterward. P10 framed a similar conditionality in explicitly moral terms, describing his own trust as resting on the founder's character rather than any check he had personally carried out:

\begin{quote}
\textit{``I trusted his (a public face of Kahf) words, his life philosophy. He is a writer, and he seemed like a good Muslim to me. Because of this I did not pay much attention. Considering his whole background, it seemed to me that this person would not ruin our privacy. Now, people can make mistakes, they will, but it seemed to me that he would not.''} (P10)
\end{quote}

P11 extended this conditional logic to potential future competitors:

\begin{quote}
\textit{``If any non-Muslim community makes such a browser, I would say the user needs to be cautious. If both are on the same page, working with the same vision in mind, that is fine. But if the agenda is different, that organization or initiative can be harmful. We would, within our capacity, make people aware that there is a certain checklist, certain criteria, they should check.''} (P11)
\end{quote}

Participants who started from religious branding often moved toward verification over time, while those who started from suspicion still left room for trust to grow once specific evidence appeared, a shared willingness to keep revising the judgment rather than treating either faith or skepticism as final.

\subsection{The Usability-Faith Bargain}

\subsubsection{Accepting Friction in Exchange for Value Alignment}

Participants often compared Kahf to faster more polished mainstream browsers and a clear pattern emerged where they tolerated the friction not out of passivity but as a deliberate choice. P01's account shows how much frustration participants absorbed before giving up entirely:

\begin{quote}
\textit{``I installed that browser at least ten to twelve times and deleted it again each time since 2024. I always keep it handy. If they bring out a good update, I will use it. Finally, the new update is something I was hoping for, though there is still scope to improve.''} (P01)
\end{quote}

Several participants stated this trade-off as a rule they applied consciously, not a compromise they drifted into. P06 set his threshold plainly, as long as the browser meets his religious requirements, a bit of slowness is not a problem. P07 went further, saying he would use a less developed or slower Muslim browser as long as it matched his values, treating alignment as more important than polish. P10 was even more direct and said he would accept any technical trade-off without any doubt if it matches his values and requirements. P05 said the same in practice, continuing to use Kahf for religious reasons despite it being less smooth than Google Chrome. P11 summarized the same compromise as:

\begin{quote}
\textit{``I do not think Kahf Browser to be the fastest, most private, or most secure browser, but rather to be the most faith-oriented, personalized browser.''} (P11)
\end{quote}

P05, P06, P07, P10 and several other participants reported that the browser's speed was comparatively slow. One reason why this bargain felt manageable was the browser's familiar look, though usage did not always feel as smooth. P13 said Kahf's interface resembles Chrome and is therefore familiar to use. P15 similarly added

\begin{quote}
\textit{``The interface is almost the same, though the usage did not feel as smooth. I used to use Chrome before, and Firefox too. Compared to Chrome, it is almost identical. But a  disadvantage I would mention that the speed feels a bit slow when loading something, it takes a bit more time compared to Google Chrome.''} (P15)
\end{quote}

When asked if this resemblance was deliberate, he agreed, explaining that developers had designed it that way on purpose and kept several features simple for the same reason. This familiarity worked as a hidden usability asset, lowering the friction of adoption. He later described dropping Facebook and YouTube from his phone entirely and shifting both personal and office work onto Kahf, showing his tolerance for rough edges was matched by real intention to keep using it.

P11 traced a specific trajectory of improvement, describing how early filtering produced frequent false positives, blurring almost everything, before recent updates substantially improved accuracy, tolerance was not indefinite, and participants expected the browser to keep closing the gap with mainstream alternatives.

\subsubsection{Where Tolerance Ends}

A sharply different pattern marks \textit{the limit of tolerance}, the point beyond which participants stopped excusing flaws. This limit tracked whether a failure touched the browser's core protective function rather than the severity of inconvenience itself: a slow page load was an annoyance, but filtering that failed to filter was a breach of the browser's basic purpose. P13's account shows exactly where that line falls:

\begin{quote}
\textit{``Kahf Browser protects you from images, various pop-ups, and harmful websites, but when it comes to videos, its performance is not very good. It can apply blurring to the thumbnail, but accidentally, once the actual video starts playing, it does not do it anymore. But it was not supposed to be like this right?''} (P13)
\end{quote}

He praised the browser's image and search filtering without reservation, but treated the gap in video filtering as unacceptable rather than a minor shortfall. P08 identified a related but distinct problem, \textit{filtering that backfires}:

\begin{quote}
\textit{``I searched for a few things myself, and it mostly blurs a lot of things. But sometimes, I felt like it draws more attention to what is in the video because of that blur effect.''} (P08)
\end{quote}

The mechanism meant to conceal content instead drew additional attention to it. She wanted content to arrive already filtered, arguing this was the only way to give a genuinely Islamic feed its intended meaning, rather than showing users the very thing it was supposed to protect them from. P10 held Kahf to a similarly strict standard, describing the platform's shortcomings as a broken promise.

\begin{quote}

\textit{``Their two main promises were that they would maintain security, and haram content would be handled properly, meaning it would stay blurred. Personally, I think the purpose is not always fulfilled and they are not fully keeping their words. The blurring works well for images, but it's not always great for videos. Overall, I would be much more satisfied if such content didn't show at all.''} (P10)
\end{quote}

He treated the browser's stated commitments in near-contractual terms, expecting both to be honored without exception. P07 proposed stricter, content-aware blocking aimed at a religious user base, while recognizing a practical risk in that same fix, he might be reading an article or a foreign book with an adult cover, and if it were fully covered, he wouldn't be able to tell if it was the original book or not. He suggested this could be advanced further by basing it on what a person is actually searching for, which would be a good moral standpoint to take.

Together, these two patterns show that the usability-faith bargain was never unconditional. Participants drew a fairly consistent line between flaws that just slowed them down and flaws that exposed them to the very content Kahf was supposed to block. The first they could excuse as a startup still finding its feet. The second, for several participants, called the browser's basic promise into question.

\subsection{Representation, Ownership, and Digital Autonomy}

\subsubsection{Pride, Representation, and Structural Imbalance}

Several participants chose Kahf not for a specific feature but for who built it. This pride in representation came up again and again, alongside religious values, in almost every interview. P01 was already looking for an alternative to Google over data-collection concerns, and what drew him to Kahf was simply that it was built by Muslims, made in Bangladesh. He switched before even testing how well it worked, choosing it for its identity rather than any proof of what it could do. P08 felt a similar pull, though hers centered on the people behind the product rather than the identity in the abstract.

\begin{quote}
\textit{``I was interested to see how the workplace was run. Since this was built by a Muslim and Bangladeshi from your own corporate world, I had a curious desire to see how it goes. When someone from your own nationality and community does something, there is a high expectation. One feels pride in it.''} (P08)
\end{quote}

P08's curiosity centered on the workplace and the people behind it, not the product itself. Her pride attached to her community succeeding on unfamiliar ground, with the browser almost incidental to it. This pride rarely stood alone, participants framed it against a sense that Muslims had been largely absent from producing technology rather than merely consuming it. P06 put this bluntly, arguing non-Muslims are ahead in technology and Muslims remain behind. For him, Kahf mattered less as a product launch than as a rare exception to a settled pattern. P10 echoed this diagnosis more concisely, framing Kahf's arrival as evidence Muslims were finally coming forward. P12 reached a similar conclusion through personal reflection:

\begin{quote}
\textit{``We need such technology very much. As a student, I had been thinking for a long time about why nothing like this was progressing, or being developed specifically for the Muslim community, or as something Muslim-based.''} (P12)
\end{quote}

P12's phrasing shows her interest predates Kahf Browser itself. Reading alongside P06's and P10's accounts, participants brought existing expectations and disappointments to their encounter with the browser, with their pride responding to a long-standing absence rather than the product's merits alone. P10 closed this line by turning it into an obligation rather than a private feeling:

\begin{quote}
\textit{``As Muslims, we should value the sacrifice and effort of our Muslim brothers, and spread this, and discuss it with friends, at the office, with family. Discussing it is what spreads it.''} (P10)
\end{quote}

\subsubsection{Platform Power, Dependency, and Governance}

For several participants, this imbalance extended into concrete concerns about where user data and advertising revenue end up, a pattern which is \textit{platform power and data autonomy}. P05 held a strong belief that mainstream platforms channel advertising revenue toward causes hostile to Muslim communities:

\begin{quote}
\textit{``He (a public face of Kahf) said in a podcast that our daily use of mainstream browsers contributes significant amounts to Google AdSense. Google then uses that money from AdSense, and part of it goes toward anti-Islamic causes.''} (P05)
\end{quote}

P05 attributed this claim to a podcast rather than to anything he had verified himself; the claim was not independently confirmed in this study. The claim still matters as data that shows a pattern of reasoning by informal citation recurring across several participants, and how mistrust of mainstream platforms became bound up for some with a religious and geopolitical frame rather than a technical one. P03 raised a similar concern in national terms, arguing that browsers built for Muslim-majority countries would keep information and resources inside those countries rather than flowing out to others.

P05 developed this further, describing his interest in an alternative platform in terms of freedom to speak without penalty, with his own community's platform. P13 gave the most developed account of this imbalance, framing dependence on major technology companies as \textit{neo-colonial dependency} through a historical lens:

\begin{quote}
\textit{``There is a concept called neo-colonialism. Large tech firms now control the economic and social technology ecosystems of most of the world. Just as colonized people were helpless under colonial powers, people today are helpless before these big-tech companies.''} (P13)
\end{quote}

What distinguishes P13's framing is not the historical comparison alone, several participants gestured toward that less directly, but his use of the comparison to explain Kahf's technical limitations rather than excuse them. He treated the browser's incompleteness as something expected rather than disqualifying, arguing a project's worth does not depend on being finished or fully competitive, comparing it to a broken shield that still lets a person do something rather than nothing, and to being Captain America rather than Iron Man, a hero whose shield is ordinary but who stands his ground with it regardless. For P13, imperfection is not merely tolerable in resistance, it is expected of such new technologies.

Not all participants treated Kahf's autonomy as already achieved. P10 pointed to a more structural barrier:

\begin{quote}
\textit{``We are not there yet. To build a complete Muslim Internet, obstacles will arise, one of them being infrastructure cost. There are also factors like geopolitics, so overall it will be expensive.''} (P10)
\end{quote}

P11 named a specific unresolved \textit{acknowledged dependency} at the project's core, one no amount of admiration could fully resolve: positioning itself against major technology companies, Kahf users still rely on infrastructure those companies control.

\begin{quote}
\textit{``Even to this day, there is no such thing as a flawless Islamic initiative anywhere in the world. Despite Kahf Browser, we are still bound to use other platforms. Even when I speak against Google, I am not exactly speaking against Google, I am speaking against Google's policy.''} (P11)
\end{quote}

P11 distinguishes opposing a company's infrastructure from opposing its policies, letting him keep criticizing major platforms without treating Kahf's technical reliance on them as a contradiction, a position more measured than the uncomplicated pride described earlier, and than P13's historical framing, showing participants could hold a critical, qualified view of Kahf's autonomy while remaining committed users of the product.

Autonomy, in these accounts, was never something Kahf simply achieved by existing. Participants weighed pride against cost, admiration against acknowledged limits, and enthusiasm against disappointment at finding the project less open than expected. Autonomy comes across less like something Kahf has, and more like something it's working toward.

\subsection{Segregation Without Isolation: Negotiating the Boundaries of a Religious Internet}

\subsubsection{Visions of Separation and Its Risks}

Participants held varied, sometimes conflicting, positions on what a separate Muslim Internet should look like. Some accepted a fully distinct system despite its costs. Others saw serious risks and preferred a more open, values-aligned alternative not restricted by religion. Some showed clear \textit{support for hard separation}, treating it as desirable and achievable. P02 held this view without hedging:
\begin{quote}
\textit{`` A different Internet system should exist, and I think it is necessary for Muslims.''} (P02)
\end{quote}

He accepted a real technical cost, framing the trade-off as communal loyalty rather than convenience, a slower Internet would simply have to be accepted. P05 reached a similar conclusion, emphasizing what community control over a platform's rules would let him do.

\begin{quote}
\textit{``The good side is this: if we had our own platform, like Facebook, and I wanted to say something against Facebook's community standards but in line with my Islamic values, I cannot say it, because Facebook or YouTube gives a strike. This could have been a good side.''} (P05)
\end{quote}

P15 supported separation for a related reason, wanting a platform where Muslims could escape the fitna he saw spread across the mainstream Internet, but he built openness into that vision from the outset rather than treating it as something to resolve later:

\begin{quote}
\textit{``I think they should definitely be allowed in. What I mainly want is a halal system for Muslims, since there's so much fitna on the Internet right now, harmful websites, and social media content that doesn't align with Islam. If all of that could be separated out, if we had our own platform entirely, that would be a good thing. And if it's our community's platform, people of other religions could still come in if they wanted. There wouldn't need to be strict rules keeping them out.''} (P15)
\end{quote}

P10 arrived at the same openness through a different route, framing it not as a concession but as an obvious, near-universal benefit:

\begin{quote}
\textit{``If anyone judges this impartially, I think they would say Kahf is necessary for Hindus, Muslims, Buddhists, Christians, everyone. In this era, if you want to maintain ethics and morality, and keep your family members safe, of course you would need this kind of thing.''} (P10)
\end{quote}

P15 and P10 both arrived at openness while supporting separation itself; other participants arrived at the same conclusion from the opposite direction, worried less about who a bounded system would exclude than about what it would cost the people already inside it, a shared concern this section terms an \textit{isolation risk}. P04 worried that separation could pull Muslims further from the wider world, resolving this by favoring a values-aligned platform open to everyone, grounded in a theological claim that Islam came for all mankind, not only Muslims. P12 raised a related worry, framed around her own exposure narrowing rather than the world's:

\begin{quote}
\textit{``I believe it would be great to have a Muslim Internet environment, but sometimes I do think: would this mean we would only learn about Muslims, and not learn about the rest of the world? I think not only Muslims, but some non-Muslims, would also use it. People who want to be more productive, and live a decent and moral life, would use it too.''} (P12)
\end{quote}

P04's concern was outward-facing, centered on globalization as a societal good; P12's was inward-facing, centered on her own exposure to the world narrowing. Both resolved this the same way, imagining the system open to non-Muslims rather than abandoning the idea. P07 reached the same preference for openness from a different angle: since content like gambling damages Muslims and non-Muslims equally, a protection built against that harm should not be withheld from non-Muslim users based on religious identity, what matters is the protection's purpose, not its marketed audience.

\subsubsection{Permeability and Governance Conditions}

P10 carried this logic of openness furthest, questioning whether Kahf itself should be understood as Muslim-exclusive at all and said:

\begin{quote}
\textit{``Islam promotes peace and brotherhood, so if Kahf moves forward and we end up with a Muslim Internet in the future, it should be open for everyone, not just Muslims. If non-Muslims join in, Alhamdulillah, we can say its for greater good. We are doing something good.''} (P10)
\end{quote}

P11 reached a related position, arguing the values Kahf protects, decency and modesty chief among them, belong to no single religious community and therefore do not justify a religiously exclusive product. He paired this with caution about the opposite extreme: too constrained an environment narrows a user's exposure to the world rather than protecting them within it, it starts to sound like controlling the user base. Enthusiasm for filtering and wariness about over-restriction were not contradictions he had to sort out, he just held both at once.

Looking beyond Kahf's current state and towards a Muslim Internet infrastructure, P06 proposed a governance sequencing for any future effort of this kind and said:

\begin{quote}
\textit{``Any future effort of this kind should not be a decision of developer only. Islamic scholars and software engineers should sit together from the earliest planning stages, well before any decision is made to scale a product or open its source code.''} (P06)
\end{quote}

A related but distinct concern ran through several accounts, not whether outsiders could enter a bounded community, but whether members already inside could freely leave it when needed. P07 wanted a boundary that functioned as a default setting rather than a fixed wall:

\begin{quote}
\textit{``Say I choose to stay within this community, to stay protected. But I now need something that is a current necessity, something not in this community. In that case, I would go outside, outside that community, outside that door, and access other resources.''} (P07)
\end{quote}

A door a user can step outside of, rather than one that seals them in permanently, captures how participants imagined these boundaries functioning, not a hard partition, but a default setting reversible at will.

Support for a bounded system depended heavily on \textit{governance conditionality} rather than any abstract commitment to boundaries themselves. P13 directs it at the entire premise of a separate Muslim Internet as a coherent, achievable project, questioning it on the grounds of the sheer geographic spread of the community any such system would need to contain:

\begin{quote}
\textit{``The reason is that Muslim identity is geographically extremely broad. From Bangladesh to practically every part of the world, Muslims exist everywhere. In the context of the Internet or technology, I can perhaps filter things to protect a community and its values. But if I try to construct a China-like wall that is simply a pro-Muslim wall around the Internet, I do not think it would work at this point of globalization there must be some openness.''} (P13)
\end{quote}

Support for a bounded Muslim Internet was rarely unconditional: participants who wanted separation still wanted an exit, and participants who wanted openness still wanted some form of protection. What ultimately decided a participant's position was less the abstract appeal of boundaries and more the specific governance, permeability, and purpose attached to them.

\section{Discussion}
This section reflects on our findings on trust, usability, and boundary-setting in faith-aligned browsing, connects them to the existing literature, and discusses implications for a faith-aligned Internet infrastructure.

\subsection{From Personal Discipline to Conditional Trust}

Previous research demonstrates that individuals tend to trust unknown systems based on category membership rather than experience. The theory of ``swift trust'' explains that in cases where people do not have time or a history to develop trust through interaction, they rely on recognized roles and institutional categories \cite{meyerson1996swifttrust}. Another approach views trust as layered, dynamic and not fixed \cite{hoffbashir2015trustautomation}. This builds on trust as a willingness to accept vulnerability \cite{mayer1995integrativetrust}, as well as the principle that trust should track what a system is capable of rather than resting on blanket confidence \cite{leesee2004trust}. Our results reflect these patterns and clarify how it operates for a faith-branded product, extending the existing literature. Religious identity may function as a basis for trust at the starting point rather than a fixed baseline. Some participants extended trust immediately on the basis of the Islamic branding or scholarly endorsement of content, whereas some others treated the same branding as insufficient, arguing that a religious label signals intent rather than guarantees conduct. Participants who framed their self-restraint as the main safeguard with Kahf in a supportive role, invoked trust in an existing practice rather than transferring moral authority to the tool. Beyond this content-level endorsement, trust for several participants was mediated personally through a local scholar vouching for them directly. This is consistent with prior HCI findings on how women’s tech adoption in Pakistan is often mediated through a trusted family member who acts as an validator rather than through the user’s own evaluation of the technology \cite{ibtasam2019cousin}. 

Many of our participants considered the founder’s character rather than the technical features of the product as enough grounds to skip verification. This reasoning recalls the folk-theory logic that users use when applying algorithms that they cannot directly inspect \cite{karizat2021folktheories, eslami2015always}. As opposed to literature that shows users are inferring that an opaque system represses their identity, our participants inferred protection from a shared one, with the founder’s religious identity functioning more as a character reference than a technical claim. This runs in the opposite direction from a parallel HCI finding in Islamic content recommendation, where Bangladeshi Muslim users reasoned that platforms secretly suppressed Islamic content because algorithms were, in their view, controlled from the West \cite{jahan2026imaan}. Both studies show Muslim users building causal explanations for an opaque system's intent from religious identity rather than technical evidence. Our findings show the same reasoning extending trust rather than suspicion when the identity in question is shared rather than foreign. 

Second, we show that this trust is conditional and revocable rather than settled once granted. Several participants used identity as a starting point to leave behind, actively looking for evidence, for instance, developer’s background, that users could verify independently of religious identity, exactly the path that swift trust theory predicts once category-based trust is checked against evidence as it becomes available  \cite{meyerson1996swifttrust}. This impulse has a documented precedent, Muslim users in the United States report learning often the hard way, to distrust apps marketed to their community after reports that a popular prayer-times application sold location data to government-linked buyers \cite{razaq2024muslim}, a caution our participants seem to have carried forward. Participants were willing to extend the same conditional trust to a future non-Muslim competitor sharing Kahf's stated values, and to withdraw it from Kahf itself following a confirmed lapse.

\subsection{The Usability-Faith Bargain as a Trust Mechanism}

Our research provides a particular mechanism for a more broader claim already made in faith-sensitive design that products aligned with religious values deserve design attention proportional to what they claim to protect, a concern with a long history in HCI \cite{buie2013spirituality} that recent work confirms remains unresolved \cite{wolf2024stillnotalot}, and that intensifies when the specific values a product claims are unfamiliar to those designing around them \cite{conwill2025challenges, naqshbandi2022makingspace}. Participants tolerated slow speed and rough interface as the usual cost of a young, community-built product but drew a hard line at faults in the browser’s core protective function. A filtering mechanism that did not block or blur haram content was interpreted as a broken promise instead of an inconvenience.

This distinction is not about severity: a slow page and a failed filter are both technical shortcomings, but only one threatens the reason participants adopted Kahf in the first place. We call this the \textit{usability-faith threshold}, the point at which a flaw stops being tolerated and starts being read as a violation. This is hardly the case for Noor's account of ritual-integrity-by-design \cite{noor2026decoding}, where friction that protects the practice is preserved as a feature. Our participants did not treat friction itself as valuable, what mattered was whether a failure signaled a broken protective function rather than mere inconvenience. Organizational trust-repair research offers a further explanation for why a failed filter reads as more damaging than slowness. Violations of competence and violations of integrity call for different repairs \cite{kimdirkscooper2009trustrepair}, and a failed filter tests the integrity in a way a slow page load does not.

Our threshold also converges with a hierarchy proposed in concurrent work on Islamic content recommendation, which places authenticity at the base of a needs hierarchy, non-negotiable and prior to appropriateness or spiritual benefit, since no content should be recommended unless first verified as free from doctrinal distortion \cite{jahan2026imaan}.  Our participants conducted the same exercise in relation to a different failure mode, one that they believed to be forgivable for being slow or unpolished yet unforgivable for failing at the one function that justifies its existence. We take this as evidence that the protective function is evaluated pass-fail while everything else is a spectrum.

This disparity is not exclusive to Kahf or Islamic contexts. Video-mediated tools interrupted the embodied rhythm of Buddhist chanting during COVID-19 for reasons structurally similar to why sometimes filtering failure breaks Kahf's core promise \cite{claisse2023buddhism}. Furthermore, Muslim women using mainstream menstrual-tracking apps observe the same pattern, the generic design overlooks the specific practice a tool claims to support \cite{ibrahim2024ramadan, ibrahim2024islamically}. This also differs from manipulative design, where dark patterns hide intent \cite{mathur2019darkpatterns}, though even a visible choice can still harm the autonomy it subverts \cite{mathur2021darkpattern}. Kahf's limitations were visible rather than concealed, and participants responded with concrete proposals rather than abandonment.

\subsection{Pride and Dependency in Faith-Branded Infrastructure}

Our study surfaces a tension specific to infrastructure of access that prior Islamic HCI work does not fully capture, since that work has focused mainly on single applications such as prayer, scripture, or recommendation tools \cite{campbellesen2026islamicai, jahan2026imaan}. Participants took real pride in Kahf's origins, describing it as proof that Muslim communities, and Bangladesh specifically, could build technology rather than only consume it.  Participants made a distinction between opposing a company’s infrastructure and opposing its policy which allowed them to criticize platforms such as Google while also relying on them.  At the same time, they were aware that Kahf remains structurally dependent on the same companies it positions itself against, and located this dependency in concrete constraints such as infrastructure cost rather than treating it as disqualifying.

A prayer application, or a recommendation layer inside an existing platform, can be judged on its own terms. A browser cannot, since its whole operation is connecting a user to other platforms it does not own, which is why dependency became so explicit here rather than in the adjacent Islamic HCI work on narrower tools. Postcolonial computing specifically identifies this dependency, relating how design inherits uneven economic relations from where a technology is built \cite{irani2010postcolonial}. The research conducted in rural Bangladesh deals with local moral and religious reasoning as a legitimate design consideration rather than as an obstacle to design around \cite{sultana2019witchcraft, sultana2020parareligious}; the pragmatic hybridity documented there, where practitioners use mainstream mobile infrastructure while retaining independent moral authority, closely resembles how our participants reasoned about Kahf's own dependency. Kwet’s analysis of digital colonialism \cite{kwet2019digitalcolonialism}, and Couldry and Mejias’s framing of data-as-extraction \cite{couldrymejias2019datacolonialism} describes a similar form of dependency at a larger scale.

Star's account of infrastructure holds that it stays invisible until it breaks, and it is exactly at breakdown that hidden dependencies become visible \cite{starruhleder1996ecology, star1999ethnography}. Kahf’s limitations  worked the same way, every failure to filter made visible not just a bug but the fact that Kahf runs on infrastructure it does not control. Participants maintained two evaluations open at once, pride in what a small, faith-driven team accomplished alongside a clear-eyed view of the infrastructure it still relies on. This departs from postcolonial computing's usual treatment of infrastructural dependency as a condition to critique; our participants treated it instead as a condition to work within, closer to what design justice scholarship has called for in marginalized-led technology work \cite{costanzachock2020designjustice}. Autonomy, for faith-branded infrastructure, works better as a direction of travel than as a state a product either has or lacks.

\subsection{Porous Digital Segregation}

Our research presents a more nuanced perspective on the binary assumption often underlying modern discussions about Internet fragmentation of a single global Internet that either persists, or splinters into separate systems individual bounded systems \cite{splinternetifri2024, splinternetEU2022}. Mueller's own position sits closer to our participants than to that binary, arguing that fragmentation itself is largely illusory and that the real danger lies in states forcing digital communication to align with territorial borders \cite{mueller2017fragment}. The participants rejected both extremes. Those in favor of creating a separate Muslim Internet still wanted an option to leave it when a specific need arose outside its boundaries, and those who expressed concerns regarding drawbacks of the idea still wanted an option of a protected default for Muslims.

This finding stands in instructive contrast to research on how Muslims already experience mainstream Internet infrastructure as a border in its own right. US Muslims describe Internet search as an extension of the surveillance they associate with physical border crossings, internalizing that logic into self-censorship, an involuntary, externally imposed boundary the literature terms a digital border \cite{razaq2024muslim}. Our participants' porous digital segregation inverts this pattern. Where that border is involuntary and enforced by an unaccountable power, our participants envision a chosen border, enforced by trusted governance, experienced as protection they can suspend at will. The two are not different intensities of the same phenomenon but structurally opposite relationships to control.

The existence of a bounded system depends more on the governance attached to it than on any unequivocal abstract commitment to boundaries. While all participants supported the idea of separation, they disagreed on which governance model should operate it. Those who turned down centralized, state-level control in favor of an internalized sense of moral restraint did not thereby reject separation itself, since what they rejected was the state as its governing authority, not separation as such. Several wanted governance to be sequenced intentionally. They wanted religious scholars and engineers to be part of the planning stage from the start. They did not want scholars to be brought in for approving already taken decisions. This resonates with the demand for scholars to sit on the loop, not just be consulted \cite{noor2026decoding}, a concern that has occupied Islamic authority online since the Internet first let users bypass traditionally trained scholars over questions of interpretation \cite{bunt2003islam}. The same demand arises in concurrent research on Islamic content recommendation, where the participants wanted a board of scholars from multiple schools of thought pre-certify the content, favoring a distributed panel to a single authority and an unmoderated algorithm was preferred \cite{jahan2026imaan}. It shows that two separate Bangladeshi samples, studying different products, converge on the same preference: plural, non-centralized religious authority built into governance from the start.

\begin{figure}[htbp]
    \centering
    \resizebox{0.8\columnwidth}{!}{%
    \begin{tikzpicture}[font=\sffamily,
        protectedBox/.style={rectangle, rounded corners=15pt, draw=teal!60!black, thick, fill=teal!20, minimum width=8.5cm, minimum height=6.5cm},
        openBox/.style={rectangle, rounded corners=15pt, draw=black!60, thick, left color=gray!10, right color=gray!30, minimum width=7cm, minimum height=6.5cm},
        innerBox/.style={rectangle, rounded corners=8pt, draw=black!80, thick, fill=white, minimum width=3.8cm, minimum height=1.5cm, align=center, text width=3.5cm, drop shadow={opacity=0.1, shadow xshift=0.5ex, shadow yshift=-0.5ex}}]
        \node[protectedBox] (ProtectedArea) at (-6.75, 0) {};
        \node[openBox] (OpenArea) at (6.0, 0) {};
        \node[align=center, anchor=north] at (-6.75, 2.8) {\Large \textbf{Protected Faith-aligned} \\ \Large \textbf{Internet}};
        \node[innerBox] (Halal) at (-8.85, 0.5) {\textbf{Halal and}\\ \textbf{Filtered Contents}};
        \node[innerBox] (Devs) at (-4.65, 0.5) {\textbf{Muslim}\\ \textbf{Developers}};
        \node[innerBox] (Scholars) at (-8.85, -1.5) {\textbf{Muslim}\\ \textbf{Scholars}};
        \node[innerBox] (Users) at (-4.65, -1.5) {\textbf{Users}};
        \node[align=center] at (6.0, 0) {\LARGE \textbf{Open Internet} \\[0.8cm] \Large Mainstream Platforms \\ \Large and Contents};
        \draw[loosely dotted, line width=2pt] (0, 3.5) -- (0, 1.2);
        \draw[loosely dotted, line width=2pt] (0, -1.9) -- (0, -3.5);
        \draw[line width=3.5pt, line cap=round, rounded corners=1.5pt]
            (-0.6, 0.8) -- (0.6, 0.8) -- (0.6, -0.8) -- (-0.6, -0.8);
        \draw[line width=2pt, fill=gray!20] (-0.6, -0.8) rectangle (0.2, 0.8);
        \draw[line width=2pt, fill=white, join=round]
            (-0.6, -0.8) -- (0.3, -1.0) -- (0.3, 1.0) -- (-0.6, 0.8) -- cycle;
        \draw[line width=2pt, line cap=round] (-0.6, -0.8) -- (-0.6, 0.8);
        \fill[black] (0.15, 0) circle (2pt);
        \node[align=center, font=\large\bfseries] at (0, -1.5) {Selective Exit};
    \end{tikzpicture}%
    }
    \caption{Porous digital segregation: A protected, faith-aligned space with a selective exit into the open Internet.}
    \Description{"Porous digital segregation: A protected, faith-aligned space with a selective exit into the open Internet." The diagram illustrates a conceptual model of two contrasting digital environments separated by a vertical dotted line. On the left is a blue box representing the "Protected Faith-aligned Internet," which houses four distinct sub-components: "Halal and Filtered Contents," "Muslim Developers," "Muslim Scholars," and "Users." In the center, spanning the dotted dividing line, is an icon of a partially open door labeled "Selective Exit," acting as a gateway between the two spaces. This door leads to the right side of the diagram, which features a grey box representing the "Open Internet," designated as the space for "Mainstream Platforms and Contents".}
    \label{fig:porous}
\end{figure}
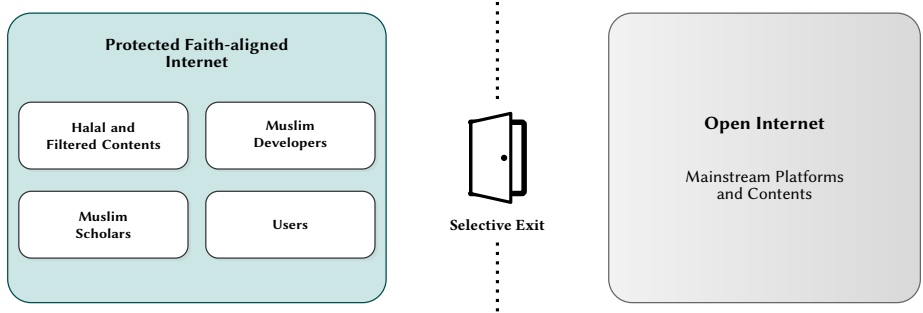

We conceptualize these relationships as \textit{porous digital segregation}, illustrated in ~\hyperref[fig:porous]{Figure 2}: a protected, faith-aligned environment whose boundary remains permeable through user-controlled selective exit. This echoes an older observation about how people respond to dissatisfaction inside a bounded group, leaving, speaking up, or staying quiet \cite{hirschman1970exitvoice}. Most platforms offer easy exit but little real voice \cite{freyschneider2023effectivevoice}, and studies of de-platforming show forced exit sends banned communities to less regulated spaces rather than reforming them \cite{chandrasekharan2017redditban, jhaver2021deplatforming}. In our study, the participants do not want such a door that isolates them and the one that would de-platform non-Muslims since they believe that all religions teach modesty so that a person can live a decent life \cite{jhaverfreyzhang2023decentralizing}. Instead, they seek a door through which provides selective exit that they control themselves. rejecting both unmoderated openness and centralized state control. Gillespie's account of moderation staying invisible to those it affects \cite{gillespie2018custodians} explains why participants wanted future governance to stay visible and accountable. We name this blueprint \textit{porous digital segregation}, a protected default users can leave whenever they like rather than a sealed alternative Internet. Participants wanted both protection and access to the wider world, with the boundary theirs to cross rather than the system's to enforce. Their worry about narrowed exposure echoes longstanding concerns about personalized filtering narrowing what a person sees without their choosing it \cite{pariser2011filterbubble, sunstein2018republic}, though they wanted that narrowing visible and reversible, which is exactly what a selective exit provides. Their language around the Ummah echoes an older account of how large communities hold together through shared belonging rather than direct contact \cite{anderson1983imaginedcommunities}.

\subsection{Design Implications}

These findings have important implications for how all protective features should be designed, how trust should be earned and how any wider faith-aligned system should be bounded.

(i) Designers should treat a product's core protective functions as categorically different from general usability. The participants used a twofold standard: lenient for polish, and strict for anything that might interfere with the browser’s protective purpose.  Attributes that define a product’s moral claim require a reliability standard closer to that employed for safety-critical than convenience, in keeping with the non-negotiable.

(ii) Transparency about technical architecture should not be optional simply because a product carries a faith-aligned identity. Many participants relied on branding alone to extend their trust. Others wanted proof they could scrutinize: what data is collected, whether there’s a back-end, and what happens to information stored locally. A faith-aligned identity may reduce the distance to initial trust, but it does not replace the transparency that a skeptical user needs to independently verify that trust.

(iii) Any Muslim Internet beyond a single browser should be designed for selective exit from the start which facilitates a visible, low-friction way to step outside a filtered environment, rather than requiring users to switch products entirely out of necessity. A system that silently shuts its own door runs the risk of recreating a narrowing of exposure which participants preferred to avoid and converge with the sort of not chosen digital border which Muslim users are already navigating elsewhere on the Internet.

\section{Limitations}
There are a few limitations of our study which we acknowledge. All participants to the study were recruited in Bangladesh and their accounts reflect a national and linguistic context. Our sample size is 16 and we note that this sample size is small, however, we noted in our methodology that this community has conservative norms. While we requested referrals from each participant, none were willing to put us in touch with anyone, thus excluding snowball sampling and limiting our ability to further grow the sample. This sample does not represent Kahf’s entire user population or a broader sample of Muslim Internet users. Kahf’s technical architectural features, comprising backend design and data flows, rely on participants’ self-reports, which were not independently verified in every instance. This study does not make any independent claims about Kahf’s actual technical architecture, our focus throughout is on how participants perceive, reason about, and trust the browser, not on auditing the browser itself. Any gap between a participant's account and Kahf's actual design is itself a finding about how users reason under conditions of limited visibility, not a claim we make about Kahf's real architecture.

\section{Conclusion and Future Work}
This study examines Kahf Browser as a lens on how Muslim users adopt, trust, and tolerate an infrastructure of access that mediates their entire relationship with the Internet. Religious identity opens the door to adoption but does not keep users inside it always, trust is revised continuously against evidence. Technical flaws are forgiven or condemned depending on whether they touch the browser's core protective promise. Pride in Muslim-built technology sits alongside a clear-eyed view of its ongoing dependency on the infrastructure it positions itself against. Our primary contribution, porous digital segregation, depicts the construction of a Muslim Internet with a protected default and selective exit, not a sealed alternative. This offers fragmentation debates a user-grounded alternative to the choice between one global Internet and a fragmented one, and gives faith-sensitive design a working threshold for which flaws a religious user base will forgive.

Our sample was drawn from a single national context and a single product, so these patterns may not generalize to other faith-aligned technologies or communities. Future work should test whether porous digital segregation holds for infrastructure that demands deeper trust, such as payment systems or app stores, whether pride in Muslim-built technology survives sustained exposure to its own dependency over time and whether this pattern reflects something specific to Islamic digital practice or a broader strategy marginalized communities use to negotiate protection against isolation.

\bibliographystyle{ACM-Reference-Format}
\bibliography{references}

\appendix

\end{document}